\documentclass[11pt, reqno]{amsproc}
\usepackage{epstopdf}
\usepackage{mathrsfs, color}
\usepackage{stmaryrd}
\usepackage{cases}
\usepackage{amssymb}
\usepackage{amsmath}
\usepackage{amsfonts}
\usepackage{graphicx}
\usepackage{amsmath,amstext,amsbsy,amssymb}
\usepackage{verbatim}
\usepackage{float}

\newtheorem{theorem}{Theorem}[section]

\theoremstyle{definition}

\theoremstyle{remark}

\numberwithin{equation}{section} \errorcontextlines=0

\begin{document}
\title[On super quantum discord for high-dimensional bipartite state]
{On super quantum discord for high-dimensional bipartite state}
\author{Jianming Zhou}
\address{Zhou: Department of Mathematics, Shanghai University, Shanghai 200444, China}
\email{272410225@qq.com}

\author{Xiaoli Hu}
\address{Hu: School of Artificial Intelligence, Jianghan University, Wuhan, Hubei 430056, China}
\email{xiaolihumath@jhun.edu.cn}

\author{Naihuan Jing*}
\address{Jing: Department of Mathematics,
   North Carolina State University,
   Raleigh, NC 27695, USA}
\email{jing@ncsu.edu}
\thanks{*Corresponding author: jing@ncsu.edu}

\begin{abstract} By quantifying the difference between quantum mutual information through weak measurement performed on a subsystem one is
led to the notion of super quantum discord. The super version is also known to be difficult to compute as the quantum discord which was captured by the projective (strong) measurements. In this paper,  we give effective bounds of the super quantum discord with or without phase damping channels for higher-dimensional bipartite quantum states, and found that the super version is always larger than the usual quantum discord as in the 2-dimensional case.
\end{abstract}
\maketitle
\section{Introduction}
The {\it quantum discord} was introduced by Olliver and Zurek \cite{OZ} to characterize the difference between the quantum mutual information before and after taking certain projective measurement on one part of the bipartite system. It is known that the quantum discord may capture quantum correlation for mixed states that goes beyond the entanglement.

Since quantum states are vulnerable to quantum measurements, the coherence of quantum states will be lost when the quantum states are measured by projective operators. However, the system may not lose its coherence completely if we perform measurements that couple the system and the measuring device weakly. For this reason, Aharonov-Albert-Vaidman proposed using weak measurement \cite{AA} in the discussion. The {\it super quantum discord} induced by the weak measurement was studied by Singh and Pati \cite{SP} and it turns out to be larger than the normal quantum discord captured by the projective (strong) measurement. In fact,
the super quantum discord covers all the values between the mutual information and the normal quantum discord. Furthermore, the super version can be potentially a more useful resource and brings new hope for further study of quantum correlations \cite{AA}.
 There are extensive studies on quantum discord and super quantum discord from different angles, most aimed at the super quantum discord for two qubit states \cite{AMC,BMR,IMS,FBP,WZL,JY,LM,WM,RSB,LC,LT,ZHJ,ZH,SM}. As high-dimension bipartite states involved a lot of more parameters, it is a difficult problem to find an exact solution of the super quantum discord.

 On the other hand, there is also an issue to formulate higher-dimensional weak measurements. Ideally one would like to define weak measurement to be a set of mutually orthogonal operators $P_i$, $i=1, \ldots, d$ such that
 (1) $P_iP_j=0$ $(i\neq j)$, (2) $P_i^2=I+\varepsilon Q_i$ for a bounded operator $Q_i$ with a parameter $\varepsilon$.
The quadratic equation satisfied by a weak projector resembles the Hecke algebra relation in representation theory, which is certain deformation of the rotation operator. Note that a general rotation in high-dimension space can be realized as a product of plane rotations with vertical axes. Along this idea, one can consider a set of weak measurements on high-dimensional Hilbert space by focusing on deforming only three projectors. We explore this idea and introduce a special set of weak measurements to study the problem.

In this paper, we study {\it an upper bound} of the super quantum discord for two classes of high-dimensional bipartite states with maximally mixed marginal. We choose three types of generators of Lie algebra $\mathfrak{su}(d)$ to construct a family of unitary operators, from which we construct a set of weak measurements. By considering an upper bound of the classical correlation on this set, we obtain an upper bound of the super quantum discord for high-dimensional bipartite states. In addition, we compare the total correlation, classical correlation and super quantum discord in high-dimensional bipartite systems. Finally, the dynamical super quantum discord for the bipartite states under phase damping channel was discussed.

 Explicitly we propose the following notion of the special weak measurements. Let $\{\Pi_0, \Pi_1, \cdots \Pi_{d-1}\}$ to be a set of orthogonal projectors such that $\Pi_i\Pi_j=\delta_{ij}\Pi_i$ and $\sum_i\Pi_i=I$.
We introduce the special weak measurement as a set of bounded operators $\{P_i(x)\}$ such that the first two operators satisfy the 2-dimensional weak measurement condition\cite{OB}. Namely,
\begin{align}\nonumber
&P_i(x)=\sqrt{\frac{1-(-1)^i\tanh{x}}{2}}\Pi_0+\sqrt{\frac{1+(-1)^i\tanh{x}}{2}}\Pi_1,\ \ \hbox{for} \ \  i=0,1,\\
&P_j(x)=\Pi_j, \ \ \hbox{for} \ \ j=3, \ldots, d-1,
\end{align}
where $x$ represents the strength of the weak measurement. In this case
\begin{equation}
\begin{split}\label{1.1}
&\hbox{(i)}\ \ \sum_i{P_i(x)}^2=I;\\
&\hbox{(ii)}\ \  [P_i(x),P_j(x)]=0;\\
&\hbox{(iii)} \ \ \lim_{x\rightarrow\infty}P_0(x)=\Pi_1, \ \  \lim_{x\rightarrow\infty}P_1(x)=\Pi_0.
\end{split}
\end{equation}

With the weak measurements $\{P_i^A(x)\}$ on subsystem $A$, the reduced density operator $\rho_i$ for a bipartite state $\rho^{AB}$ is given by
\begin{equation}\label{1.2}
\begin{split}
\rho_i=\frac{1}{p_i}\mathrm{Tr}_A{[(P_i^A(x) \otimes I)\rho^{AB}(P_i^A(x) \otimes I)]},
\end{split}
\end{equation}
where $p_i=\mathrm{Tr}[(P_i^A(x) \otimes I)\rho^{AB}(P_i^A(x) \otimes I)]$ is the  probability of the outcome $\textit{i}$.
Then the variant quantum mutual information can be defined by \cite{P}
\begin{equation}
\begin{split}
\mathcal{I}(\rho^{AB}) = S(\rho^A) + S(\rho^B) - S(\rho^{AB}),
\end{split}
\end{equation}
where $S(\rho^X)$ stands for the von Neumann entropy of the quantum state on system X.
And the classical information is given by \cite{HV}
\begin{equation}
\begin{split}
\mathcal{J}(\rho^{AB}) = \sup_{\{P_i^A(x)\}}\mathcal{I}(\rho^{AB}|\{P_i^A(x)\}),
\end{split}
\end{equation}
where $\mathcal{I}(\rho^{AB}|\{P_i^A(x)\})=S(\rho^B) - S(\rho^{AB}|\{P_i^A(x)\})$ is the variant of quantum mutual information (with respect to weak measurements $\{P_i^A(x)\}$) and $S(\rho^{AB}|\{P_i^A(x)\})= \sum_ip_iS(\rho_i)$ is the quantum conditional entropy.

The so called the {\it super quantum discord} of a bipartite quantum state $\rho^{AB}$ with the local weak measurement $\{P_i^A(x)\}$ on subsystem $A$ is the difference between the mutual information $\mathcal{I}(\rho^{AB})$ and the classical correlation $\mathcal{J}(\rho^{AB})$\cite{SP}, i.e.
\begin{equation}\label{SD}
\begin{split}
\mathcal{SD}(\rho^{AB})=\mathcal{I}(\rho^{AB})-\mathcal{J}(\rho^{AB}).
\end{split}
\end{equation}

This paper is organized as follows. In section II we give analytic formulas of the super quantum discord for two higher-dimensional states. We consider the relationships among the quantum mutual information $\mathcal{I}(\rho^{AB})$, the classical correlation $\mathcal{J}(\rho^{AB})$ and the super quantum discord $\mathcal{SD}(\rho^{AB})$ for high-dimensional quantum states.
In section III we give the dynamics super quantum discord under non-dissipative channels. The conclusion is given in section IV.

\section{The upper bounds of the super quantum discord for high-dimensional bipartite states}
It is well-known that projective measurements can be generated by the action of the special unitary Lie algebra $\mathfrak{su}(d)$ on a fixed canonical orthonormal set, so the set of projective measurements has $d^2-1$ real parameters on a $d$-dimensional space. In general it is difficult to calculate the super quantum discord for most quantum states given that weak measurements are more complicated. In this article, we consider some special situations of high-dimensional bipartite quantum states and compute their super quantum discords. On the space $H_A\otimes H_B=\mathbb C^d\otimes \mathbb C^d$ we consider the following quantum state:
\begin{equation}\label{2.1}
\begin{split} \rho^{AB}=\frac{1}{d^2}(I\otimes I+\sum_{i=1}^{|\mathcal{A}|}c_i\sigma_i\otimes \sigma_i),
\end{split}
\end{equation}
where $\sigma$ are elements of the prescribed basis, say the Gell-Mann basis, and $\mathcal A$ is a subset of the 
generators of $\mathfrak{su}(d)$.

Let $X=\{v_{ij},u_{ij},w_k|0\leq i< j< d, 0<k<d\}$ be the set of generators for the unitary Lie algebra $\mathfrak{su}(d)$, where
\begin{equation}
\begin{split}
u_{ij}&=|i\rangle\langle j|+|j\rangle\langle j|,\quad  v_{ij}=\sqrt{-1}(|j\rangle\langle i|-|i\rangle\langle j|),\\
w_k&=\sqrt{\frac{2}{k(k+1)}}(\sum_{i=0}^{k}|i\rangle\langle i|-k|k\rangle\langle k|).
\end{split}
\end{equation}

Select $\mathcal{A}=\{v_{01},u_{01},w_1,v_{ij},u_{ij}|1<i<j<d\}\subset X$ and name the first three generators as $\sigma_1=u_{01}, \sigma_2=v_{01}, \sigma_3=w_{1}$, and the remaining generators are denoted by $\sigma_i$, so $\mathcal{A}=\{\sigma_i|1\leq i\leq |\mathcal{A}|\}$. Note that $\sigma_i, i=1, 2, 3$ can be viewed
as the Pauli spin matrices, which generate the special Lie algebra
$\mathfrak{su}(2)$. Let $V_0=tI+\sqrt{-1}\sum_{i=1}^{3}y_i\sigma_i$ with $t,y_i\in\mathbb{R}$ and $t^2+\sum_iy_i^2=1$.

Let $\{|i\rangle,  i=0,1,\cdots, d-1\}$ be the computational basis of $H$,
and consider the weak measurement operators $\{\tilde{P}_i^A(x)\}$ on subsystem $A$ as follows.
\begin{equation}\label{2.3}
\begin{split}
&\tilde{P}_0^A(x)=\sqrt{\frac{1-\tanh{x}}{2}}V_0\Pi_0V_0^{\dag}+\sqrt{\frac{1+\tanh{x}}{2}}V_0\Pi_1V_0^{\dag};\\
&\tilde{P}_1^A(x)=\sqrt{\frac{1+\tanh{x}}{2}}V_0\Pi_0V_0^{\dag}+\sqrt{\frac{1-\tanh{x}}{2}}V_0\Pi_1V_0^{\dag};\\
&\tilde{P}_i^A(x)=V_0\Pi_iV_0^{\dag}=\Pi_i\ \ \hbox{for} \ \  2\leq i\leq d-1,
\end{split}
\end{equation}
where $\Pi_i=|i\rangle\langle i|$ for $i=0,1,\cdots, d-1$. We have the following relations
 \begin{equation}\begin{split}
 \Pi_0\sigma_3\Pi_0&=\Pi_0;  \ \ \Pi_1\sigma_3\Pi_1=-\Pi_1;\\
  \Pi_j\sigma_k\Pi_j&=0 \ \ \hbox{for} \ \ j=0,1, k=1,2;\\
  \Pi_k\sigma_l\Pi_k&=0 \ \ \hbox{for} \ \ l=4,\cdots,|\mathcal{A}|, k=2,\cdots, d-1.
\end{split} \end{equation}
It is readily seen that the projectors $\{\tilde{P}_i^A(x)\}$ satisfy \eqref{1.1}.
Recall that  $\mathcal{J}(\rho^{AB})$ is the supremum by traversing over general weak measurement.
The above weak measurement is a special one, so the supremum  $\mathcal{\tilde{J}}(\rho^{AB})$ over this one
obeys 
$\mathcal{J}(\rho^{AB})\geq \mathcal{\tilde{J}}(\rho^{AB})$. One can  directly check the following relations
\begin{equation}
\begin{split}
V_0^{\dag}\sigma_1V_0&=(t^2+y_1^2-y_2^2-y_3^2)\sigma_1+2(y_1y_2+ty_3)\sigma_2+2(y_1y_3-ty_2)\sigma_3;\\
V_0^{\dag}\sigma_2V_0&=(t^2+y_2^2-y_1^2-y_3^2)\sigma_2+2(y_2y_3+ty_1)\sigma_3+2(y_1y_2-ty_3)\sigma_1;\\
V_0^{\dag}\sigma_3V_0&=(t^2+y_3^2-y_1^2-y_2^2)\sigma_3+2(y_1y_3+ty_2)\sigma_1+2(y_2y_3-ty_1)\sigma_2;\\
V_0^{\dag}\sigma_kV_0&=\sigma_k \ \ \hbox{for} \ \ k=4,\cdots,|\mathcal{A}|.
\end{split}
\end{equation}
Introduce new variables $z_1,z_2,z_3$ by $z_1=2(y_1y_3-ty_2), z_2=2(y_2y_3+ty_1), z_3=(t^2+y_3^2-y_1^2-y_2^2)$,
then $\sum_i^3z_i^2=1$. From Eq.(\ref{1.2}), we have
\begin{equation}
\begin{split}
\rho_0&=\frac{1}{d}(I-\sum_{i=1}^3c_iz_i\sigma_i\tanh{x}); \\
\rho_1&=\frac{1}{d}(I+\sum_{i=1}^3c_iz_i\sigma_i\tanh{x});\\
\rho_k&=\frac{I}{d} \quad \hbox{for}\ \  k=2,\cdots,d-1,
\end{split}
\end{equation}
and $p_i=\frac{1}{d}$ for $i=0,\cdots,d-1$. The eigenvalues of $\rho_i$ are given by
\begin{equation}
\begin{split}
\lambda_{\rho_0}^{l}&=\lambda_{\rho_1}^{l}=\frac{1}{d}(1+(-1)^l\theta) \quad \hbox{for}\ \ l=0,1;\\
 \lambda_{\rho_0}^{j}&=\lambda_{\rho_1}^{j}=\frac{1}{d} \quad \hbox{for}\ \ j=2,\cdots,d-1;\\
\lambda_{\rho_k}^{j'}&=\frac{1}{d} \quad \hbox{for}\ \ {j'}=0,\cdots,d-1,
\end{split}
\end{equation}
where $\theta=\sqrt{c_1^2z_1^2+c_2^2z_2^2+c_3^2z_3^2}\tanh{x}$. Let $H(x)=(1+x)\log_2(1+x)+(1-x)\log_2(1-x)$ be the entropic function, then
\begin{equation}
\begin{split}
&S(\rho_0)=S(\rho_1)=\log_2d-\frac{1}{d}H(\theta);\\
&S(\rho_2)=\cdots=S(\rho_{d-1})=\log_2d.
\end{split}
\end{equation}
By definition 
we have $\mathcal{I}(\rho^{AB}|\{\tilde{P}_i^A(x)\})=S(\rho^{B})-S(\rho^{AB}|\{\tilde{P}_i^A(x)\})
=\frac{2}{d}H(\theta)$.

Let $F(\theta)=\frac{2}{d}H(\theta)$, then $\frac{\partial{F}}{\partial{\theta}}=\frac{2}{d}\log_2\frac{1+\theta}{1-\theta}$. Therefore, $F(\theta)$ is an increasing function with respect to $\theta\in[0,1]$. Set $c=\max{\{|c_1|,|c_2|,|c_3|\}}$, then $\theta\leq c\tanh{x}$, and the equality can be achieved by appropriate choice of $t$ and $y_i$. Hence, we can get $\max_{\{\tilde{P}_i^A(x)\}}\theta=c\tanh{x}.$ Then, 
\begin{equation}
\begin{split}
\mathcal{\tilde{J}}(\rho^{AB})=S(\rho^{B})-\min_{\{\tilde{P}_i^A(x)\}}S(\rho^{AB}|\{\tilde{P}_i^A(x)\})=\frac{2}{d}H(c\tanh{x}).
\end{split}
\end{equation}
Note that the overall minimum for all the weak measurements
is less than the minimum over this special one. Hence, we have the following result.
\begin{theorem}\label{Th1}
The upper bound of the super quantum discord for states in \eqref{2.1} is given by
\begin{equation}\label{2.10}
\begin{split}
\mathcal{SD}(\rho^{AB})&=\mathcal{I}(\rho^{AB})-\mathcal{J}(\rho^{AB})\\
&\leq\mathcal{I}(\rho^{AB})-\mathcal{\tilde{J}}(\rho^{AB})
=\mathcal{I}(\rho^{AB})-\frac{2}{d}H(c\tanh{x}).
\end{split}
\end{equation}
\end{theorem}

Note that $H(x)$ is an even function and $\frac{\mathrm{d}H}{\mathrm{d}x}\geq 0$ when $x\in(0,\infty)$, so $H(x)$ is strictly monotonically decreasing in $(-\infty,0)$ and strictly monotonically increasing in $(0,\infty)$. Therefore, the right hand side of \eqref{2.10} decreases as $|x|$ increases, and when $|x|\rightarrow {\infty}$, the right hand side of \eqref{2.10} matches with the results in \cite{HJW}. That is to say, the upper bound of the super quantum discord for the quantum state in \eqref{2.1} is larger than the quantum discord (defined by projective measurements).

Next we are going to consider the following quantum state with maximally mixed marginals:
\begin{equation}\label{2.11}
\begin{split}
\rho^{AB}&=\frac{1}{d^2}(I\otimes I+\sum_{j=1}^{\mathcal{A}}\sum_{k=1}^3t_{jk}\sigma_j\otimes \sigma_k).
\end{split}
\end{equation}
Then we have
\begin{equation}
\begin{split}
\rho_0&=\frac{1}{d^2}(I-\tanh{x}\sum_{j=1}^{3}\sum_{k=1}^3t_{jk}z_j\sigma_k);\\
\rho_1&=\frac{1}{d^2}(I+\tanh{x}\sum_{j=1}^{3}\sum_{k=1}^3t_{jk}z_j\sigma_k);\\
\rho_k&=\frac{I}{d^2} \quad \hbox{for}\ \ k=2,\cdots,d-1,
\end{split}
\end{equation}
with the probability $p_i(x)=\frac{1}{d}$ for $i=0,\cdots,d-1$.

Let $\bar{\theta}=\sqrt{\sum_{j=1}^3(\sum_{i=1}^3t_{ij}z_i)^2}\tanh{x}$. If follows from definition that
\begin{equation}
\begin{split}
\mathcal{I}(\rho^{AB}|\{\tilde{P}_i^A(x)\})=S(\rho^{B})-S(\rho^{AB}|\{\tilde{P}_i^A(x)\})=
\frac{2}{d}H(\bar{\theta}).
\end{split}
\end{equation}

To compute the super quantum discord, let's estimate the minimum of $S(\rho^{AB}|\{\tilde{P}_i^A(x)\})$.
For this we claim that
$$\max_{\{p_i^A(x)\}}\bar{\theta}\geq \bar{t}\tanh{x},$$
where $\bar{t}=\max{\{\sqrt{\sum_{j=1}^3t_{1j}^2},\sqrt{\sum_{j=1}^3t_{2j}^2},\sqrt{\sum_{j=1}^3t_{3j}^2}\}}$.

In fact, choose
$|t|=|y_1|=\frac{\sqrt{2}}{2}$, $y_2=y_3=0$, i.e., $|z_2| = 1$, $z_1=z_3=0$, then $\bar{\theta}=\sqrt{\sum_{j=1}^3t_{2j}^2}\tanh{x}$.
Similarly choose $|t|=|y_2|=\frac{\sqrt{2}}{2}$, $y_1=y_3=0$, i.e., $|z_1|=1$, $z_2=z_3=0$, then $\bar{\theta}=\sqrt{\sum_{j=1}^3t_{1j}^2}\tanh{x}$.
Then choose $|t|=|y_3|=\frac{\sqrt{2}}{2}$, $y_1=y_2=0$, i.e., $|z_3|=1$, $z_1=z_2=0$, so $\bar{\theta}=\sqrt{\sum_{j=1}^3t_{3j}^2}\tanh{x}$.
By the triangular inequality the claim holds.

Using the claim, we have that
\begin{equation}
\begin{split}
\mathcal{\tilde{J}}(\rho^{AB})=S(\rho^B)-\min_{\{\tilde{P}_i^A(x)\}}S(\rho|\{\tilde{P}_i^A(x)\})
\geq\frac{2}{d}H(\bar{t}\tanh{x}).
\end{split}
\end{equation}
Note that the weak measurement is a special one, thus we have proved the following result.
\begin{theorem}\label{Th2}
The upper bound of the super quantum discord for the state in \eqref{2.11} is given by
\begin{equation}\label{2.15}
\begin{split}
\mathcal{SD}(\rho^{AB})&=\mathcal{I}(\rho^{AB})-\mathcal{J}(\rho^{AB})\\
&\leq\mathcal{I}(\rho^{AB})-\mathcal{\tilde{J}}(\rho^{AB})
\leq\mathcal{I}(\rho^{AB})-\frac{2}{d}H(\bar{t}\tanh{x}).
\end{split}
\end{equation}
\end{theorem}

Theorem \ref{Th2} shows that the super quantum discord is larger than the quantum discord. When $x>>0$, the right in \eqref{2.15} becomes the upper bound of the quantum discord in \cite{HJW}.

In particular, we analyze the super quantum discord for the following quantum state:
\begin{equation}\label{2.16}
\begin{split}
\rho^{AB}&=\frac{1}{d^2}(I\otimes I+\sum_{j=1}^{3}c_j\sigma_j\otimes \sigma_j),
\end{split}
\end{equation}
which is a special case of Eq.(\ref{2.1}) and Eq.(\ref{2.11}). The eigenvalues of $\rho^{AB}$ are
\begin{equation}\label{2.17}
\begin{split}
&\lambda_{i}=\frac{1}{d^2}(1-c_1+(-1)^i(c_2+c_3))~ \hbox{for}~ i=0,1; \\
&\lambda_{j}=\frac{1}{d^2}(1+c_1+(-1)^j(c_2-c_3))~\hbox{for}~ j=2,3;\\
&\lambda_{4}=\lambda_{5}=\cdots=\lambda_{d-1}=\frac{1}{d^2}.
\end{split}
\end{equation}
Since the eigenvalues are nonnegative, $|c_1|+|c_2|+|c_3|\leq 1$. The reduced density operators of $\rho^{AB}$ are given by
\begin{equation}\label{2.18}
\begin{split}
\rho^A=\rho^B=\frac{I}{d}.
\end{split}
\end{equation}
Set $\tau_k=d^2\lambda_k$ for $k=0,1,2,3$, then the quantum mutual information is
\begin{equation}
\begin{split}
\mathcal{I}(\rho^{AB})=\frac{1}{d^2}\sum_{k=0}^3\tau_k\log_2\tau_k.
\end{split}
\end{equation}
The upper bound of the super quantum discord for the state in Eq.(\ref{2.16}) is
\begin{equation}\label{2.20}
\begin{split}
\mathcal{SD}(\rho^{AB})&=\mathcal{I}(\rho^{AB})-\mathcal{J}(\rho^{AB})\leq\mathcal{I}(\rho^{AB})-\mathcal{\tilde{J}}(\rho^{AB})\\
&=\frac{1}{d^2}[\sum_{k=0}^3\tau_k\log_2\tau_k-2H(c\tanh{x})],
\end{split}
\end{equation}
where $c=\max{\{|c_1|,|c_2|,|c_3|\}}$.

\begin{figure}[H]
\flushright
\includegraphics[width=1.1\linewidth]{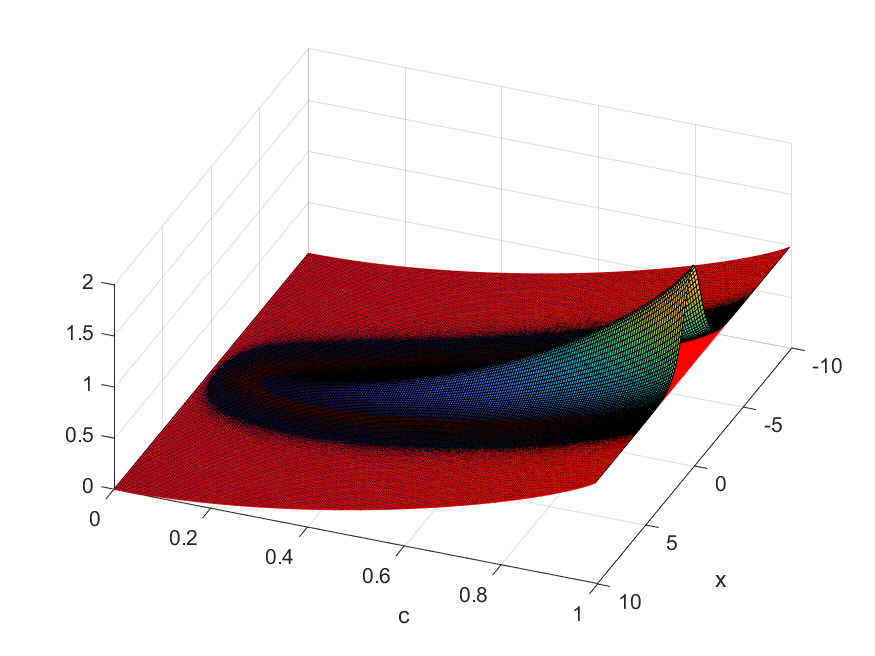}
\caption{Graphs of the super quantum discord (blue surface) and the quantum discord (red place) for state of \eqref{2.16} with $c_1=c_2=c_3=-c$ and $d=2$.}
\label{fig:1}
\end{figure}

\begin{figure}[H]
\flushright
\includegraphics[width=1.1\linewidth]{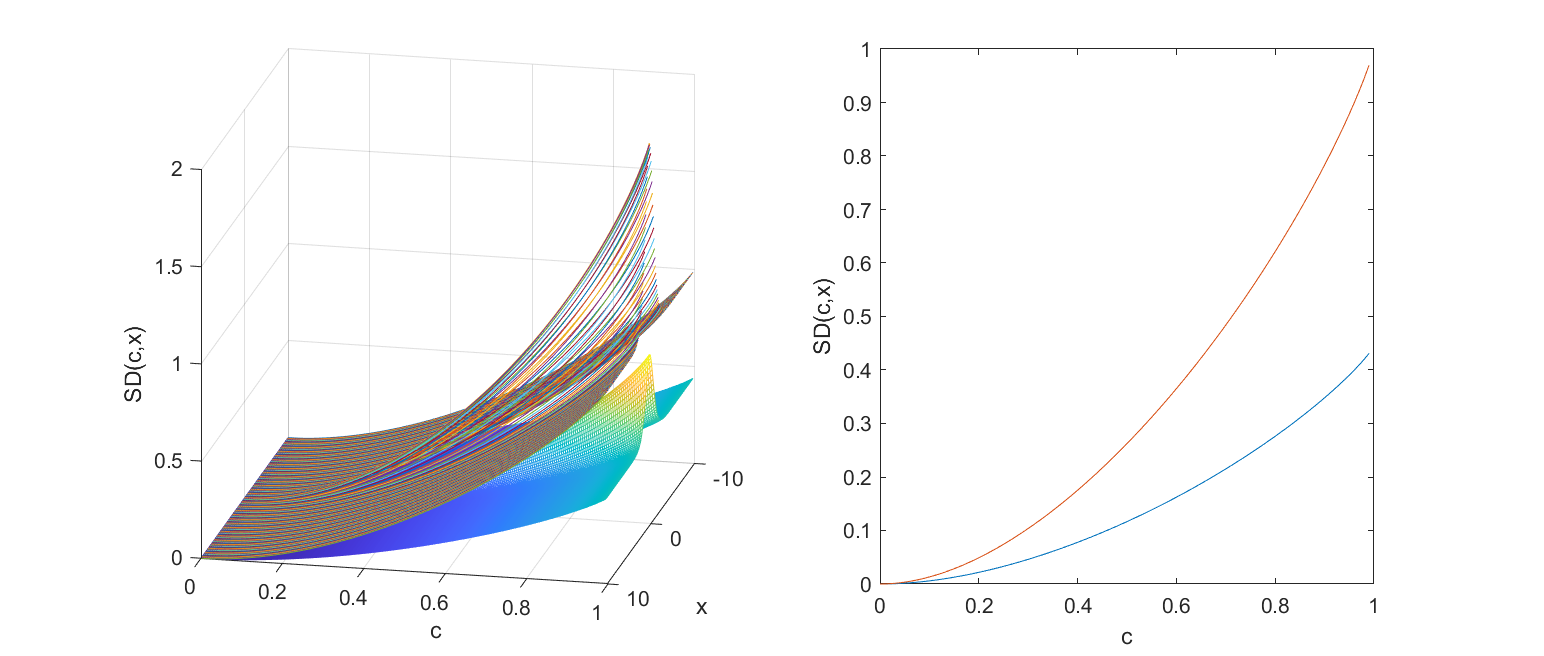}
\caption{Behavior of $\mathcal{SD}(c,x)$ for $d=3$ and $d=2$. The right graph represents the overall diagram when $d = 2$ (gray surface) and $d=3$ (blue surface), respectively. The left graph shows that our results are equivalent to those of \cite{HJW} when $x>>0$, where the blue line represents the upper bound when $d=3$, the red line represents the upper bound when $d=2$.}
\label{fig:2}
\end{figure}

\emph{Example 1.} Let $c_1=c_2=c_3=-c$ in Eq.(\ref{2.16}), where $c\in[0,1]$. According to Theorem 2.2, we have
\begin{equation}\label{2.21}
\begin{split}
&\mathcal{SD}(\rho^{AB})=\mathcal{I}(\rho^{AB})-\mathcal{J}(\rho^{AB})\leq\mathcal{I}(\rho^{AB})-\mathcal{\tilde{J}}(\rho^{AB})=\mathcal{SD}(c,x),
\end{split}
\end{equation}
where
\begin{equation}\label{2.22}
\begin{split}
\mathcal{SD}(c,x)=&\frac{1}{d^2}[3(1-c)\log_2(1-c)+(1-3c)\log_2(1-3c)
-2H(c\tanh{x})].
\end{split}
\end{equation}
The equality of \eqref{2.22} can be achieved when $d=2$.
Fig. \ref{fig:1} describes the behavior of the super quantum discord and the quantum discord at $d=2$.
It can be seen that the super quantum discord approaches the quantum discord when $|x|\rightarrow \infty$.
Meanwhile, the super quantum discord increases as $|x|\rightarrow 0$ and reaches its maximum value at $x=0$, indicating that the weak measurement is the weakest.
Fig. \ref{fig:2} shows the value distribution of the upper bound $\mathcal{SD}(c,x)$ at $d=2$ and $d=3$.
In addition, the upper bound in \eqref{2.21} is zero when $\lim d\rightarrow\infty$, which implies that the super quantum discord of the state defined by \eqref{2.16} will vanish.

For a bipartite state $\rho^{AB}$, the quantum mutual information $\mathcal{I}(\rho^{AB})$, the classical correlation $\mathcal{C}(\rho^{AB})$ and the quantum discord $\mathcal{Q}(\rho^{AB})$ are explained as the quantity of the total correlation, classical correlation and quantum correlation after measurement, respectively \cite{OZ,HV}.
Their relations have been studied in detail for high-dimensional quantum states in \cite{HJW}. In the sequel we will examine further relations among the quantum mutual information $\mathcal{I}(\rho^{AB})$, the classical correlation $\mathcal{J}(\rho^{AB})$ and the super quantum discord $\mathcal{SD}(\rho^{AB})$ for higher-dimensional quantum states.

We know that for any two-qudit pure state $\rho^{AB}=|\psi\rangle\langle\psi|$, $|\psi\rangle$ has Schmidt decomposition $|\psi\rangle = \sum_{j=0}^{d-1}a_j|j\rangle\otimes |j\rangle$ and $E = -\sum_{j=0}^{d-1}|a_j|^2\log_2|a_j|^2$ is its reduced von Neumann entropy \cite{L}. Then the reduced von Neumann entropy is equal to its supper quantum discord. i.e.,
 \begin{equation}
\begin{split}
\mathcal{I}(\rho^{AB})=2E, \ \ \mathcal{J}(\rho^{AB}) = E, \ \ \mathcal{SD}(\rho^{AB}) = E.
\end{split}
\end{equation}
Therefore, the total correlations are evenly divided into classical correlations and quantum correlations in this case.
Let's consider another extreme case in \cite{HJW}, when $\rho$ can be represented as
 \begin{equation}
\begin{split}
\rho^{AB} = \sum_{i=0}^{d-1}\sum_{j=0}^{d-1}p_{ij}|i\rangle\langle i|\otimes |j\rangle\langle j|
\end{split}
\end{equation}
with a bivariate probability distribution  $p=\{p_{ij}\}$. In this case, all correlations are classical and there are no quantum correlation, i.e.,
 \begin{equation}
\begin{split}
\mathcal{I}(\rho^{AB})=I(p), \ \ \mathcal{J}(\rho^{AB}) = I(p), \ \ \mathcal{SD}(\rho^{AB}) = 0,
\end{split}
\end{equation}
where $I(p)$ is the classical mutual information.
We know that in the quantum discord version, $\mathcal{C}(\rho^{AB})\geq\mathcal{Q}(\rho^{AB})$ hold on most case. And the above two examples also show that $\mathcal{J}(\rho^{AB})\geq\mathcal{Q}(\rho^{AB})$, but $\mathcal{J}(\rho^{AB})\geq\mathcal{Q}(\rho^{AB})$ is incorrect for all states.

We now conduct numerical analysis of the super quantum discord.
Recall that $\mathcal{J}(\rho)$ and $\mathcal{\tilde{J}}(\rho)$ are the supremums by traversing the general weak measurements $\{P_i^A(x)\}$
and the special subset $\{\tilde{P}_i^A(x)\}$ respectively. We denote by $\mathcal{D}(\rho)=\mathcal{J}(\rho)-\mathcal{SD}(\rho)$ the difference between the classical correlation and the super quantum discord.
Next we consider the state in \eqref{2.16}). As a matter of fact $\mathcal{D}(\rho)=\mathcal{J}(\rho)-\mathcal{SD}(\rho)= 2\mathcal{J}(\rho)-\mathcal{I}(\rho)\geq2\mathcal{\tilde{J}}(\rho)-\mathcal{I}(\rho)=\mathcal{D}(c_1,c_2,c_3,x)$.
Then by Eq.(\ref{2.20}), $\mathcal{D}(c_1,c_2,c_3,x)$ is also a lower bound of $\mathcal{D}(\rho)$, where
 \begin{equation}\label{2.26}
\begin{split}
\mathcal{D}(c_1,c_2,c_3,x)&=\frac{1}{d^2}[4H(c\tanh{x})
-\sum_{k=0}^3\tau_k\log_2\tau_k].
\end{split}
\end{equation}

\begin{figure}[H]
\centering
\includegraphics[width=5in]{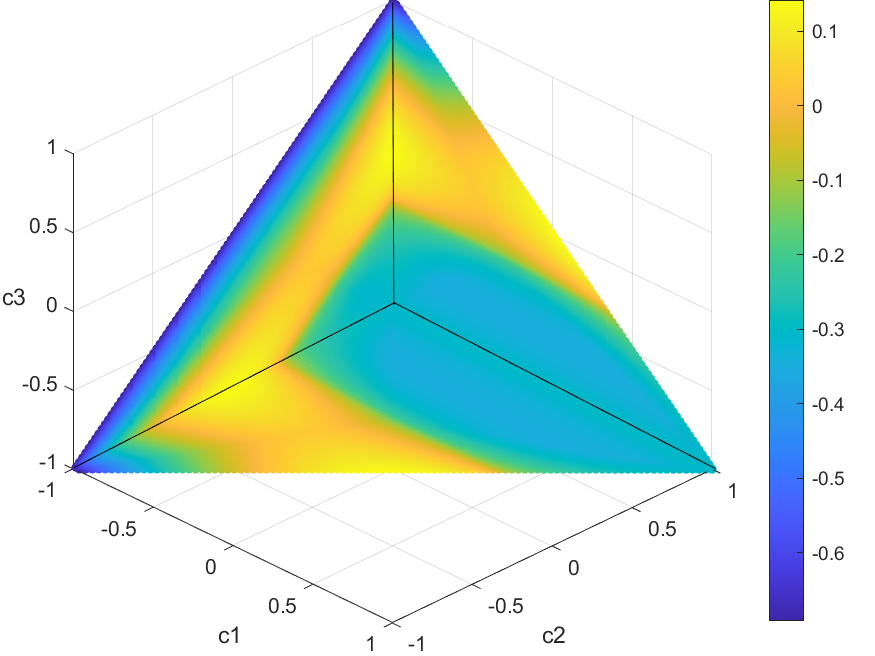}
\caption{Behavior of $\mathcal{D}(c_1,c_2,c_3,x)$ with $c_1,c_2,c_3\in [0,1]$ for $\rho^{AB}$ in Eq.(\ref{2.16}) when $x=0.5$ and $d=3$. These points are concentrated in the tetrahedron because c is limited by the non-negative eigenvalue $\lambda_i$.}
\label{fig:3}
\end{figure}

\begin{figure}[H]
\centering
\includegraphics[width=5in]{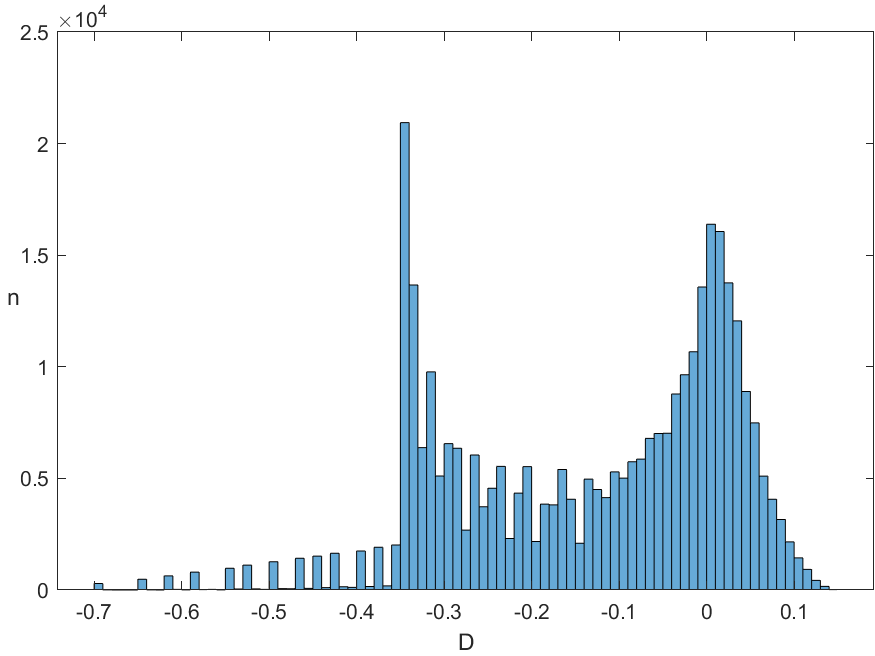}
\caption{The statistical distribution of $\mathcal{D}(c_1,c_2,c_3)$ at $d=3$ and $x = 0.5$ containing 332980 samples. The x- and y-coordinates respectively represent the values of $\mathcal{D}(c_1,c_2,c_3)$ and the number of samples distributed in the corresponding interval. About $27.66\%$ values of $\mathcal{D}(c_1,c_2,c_3)$ are non-negative.}
\label{fig:4}
\end{figure}

Fig.\ref{fig:3} shows the graph of $\mathcal{D}(c_1,c_2,c_3,x)$ when $d=3$ and $x=0.5$ for state defined by \eqref{2.16}, while its statistical distribution is depicted in Fig. \ref{fig:4}. About $27.66\%$ values of $\mathcal{D}(c_1,c_2,c_3)$ are non-negative. This value of the discord version is about $88.55\%$ in \cite{HJW}.
Moreover, $\mathcal{D}(c_1,c_2,c_3,x)$ is strictly monotonically increasing when $x\in(0,\infty)$ and independent from the dimension $d$.
Therefore, we conclude that as $|x|$ decreases, the measurement intensity weakens and
$\mathcal{J}(\rho)\leq\mathcal{SD}(\rho)$ holds for fewer high-dimensional quantum states in \eqref{2.16}.

\section{Dynamics of super quantum discord under nondissipative channels}
In this section, we will discuss the behavior of the correlations for the state in \eqref{2.1} through the phase damping channels with the Kraus operators. The Kraus operators is defined by
\begin{equation}\label{3.1}
\begin{split}
E_0=\sqrt{\gamma}I,\quad
E_1=\sqrt{1-\gamma}\left(
  \begin{array}{ccc}
    0 & 1 & 0 \\
    1 & 0 & 0 \\
    0 & 0 & I_{d-2} \\
  \end{array}
\right),
\end{split}
\end{equation}
where $\gamma\in[0,1]$ and $\sum_iE_iE_i^\dag=I$. The Kraus operaotrs characterize the evolution of super quantum discord under bit-flip noise. For two-qudit states, the action of the phase damping channel $\varepsilon$ on $\rho^{AB}$ is given by 
\begin{equation}\label{3.2}
\begin{split}
\varepsilon(\rho^{AB})=(E_0\otimes I)\rho^{AB}(E_0\otimes I)^{\dag}+(E_1\otimes I)\rho^{AB}(E_1\otimes I)^{\dag}.
\end{split}
\end{equation}
The quantum states in \eqref{2.1} under the phase damping channel are changed into
\begin{equation}\label{3.2}
\begin{split}
\tilde{\rho}^{AB}&=\varepsilon(\rho^{AB})=\frac{1}{d^2}[I\otimes I +c_1\sigma_1\otimes \sigma_1\\
&\quad +(2\gamma-1)(c_2\sigma_2\otimes \sigma_2+c_3\sigma_3\otimes \sigma_3)+\sum_{k=4}^{|\mathcal{A}|}c_k\sigma_k\otimes \sigma_k].
\end{split}
\end{equation}
By theorem \ref{Th1}, we can get the following results.
\begin{theorem}\label{thm3}
For the states in \eqref{2.1} through the phase damping channel, the upper bound of their super quantum discord is given by
\begin{equation}\label{3.3}
\begin{split}
\mathcal{SD}(\tilde{\rho}^{AB})\leq \mathcal{I}(\tilde{\rho}^{AB})-\tilde{J}(\tilde{\rho}^{AB})=\mathcal{I}(\tilde{\rho}^{AB})-\frac{2}{d}H(\bar{c}\tanh{x}).
\end{split}
\end{equation}
where $\bar{c}=\max{\{|c_1|,|(2\gamma-1)c_2|,|(2\gamma-1)c_3|\}}$.
\end{theorem}
We see that the parameter $\gamma$ also determines the degree of noise influence for the super quantum discord after the channel. Obviously, the super quantum discord remains unchanged when $\gamma = 0$. Meanwhile, the upper bound of the super quantum discord is a decreasing function with respect to $x$. Therefore, it is meaningful to observe the dynamic evolution of $\mathcal{SD}(\tilde{\rho}^{AB})$ under phase damped channel.
The state in \eqref{2.16} through the phase damping channels is changed to
\begin{equation}\label{3.4}
\begin{split}
\tilde{\rho}^{AB}=\frac{1}{d^2}[I\otimes I +c_1\sigma_1\otimes \sigma_1+(2\gamma-1)(c_2\sigma_2\otimes \sigma_2+c_3\sigma_3\otimes \sigma_3)],
\end{split}
\end{equation}
which is a special case in \eqref{3.2}. The eigenvalues of $\tilde{\rho}^{AB}$ are
\begin{equation}\label{3.5}
\begin{split}
&\lambda_{i}=\frac{1}{d^2}(1-c_1+(-1)^i(2\gamma-1)(c_2+c_3)),\ \ \hbox{for} \ \ i=0,1; \\ &\lambda_{j}=\frac{1}{d^2}(1+c_1+(-1)^j(2\gamma-1)(c_2-c_3)),\ \ \hbox{for} \ \ j=2,3;\\
&\lambda_{4}=\lambda_{5}=\cdots=\lambda_{d-1}=\frac{1}{d^2}.
\end{split}
\end{equation}
Set $\tilde{\tau}=d^2\lambda_k$ for $k=0,1,2,3$, we have
\begin{equation}\label{3.6}
\begin{split}
\mathcal{SD}(\tilde{\rho}^{AB})\leq& \frac{1}{d^2}[\sum_{k=0}^3\tilde{\tau}\log_2\tilde{\tau}
-2(1+\bar{c}\tanh{x})\log_2(1+\bar{c}\tanh{x})\\
&-2(1-\bar{c}\tanh{x})\log_2(1-\bar{c}\tanh{x})].
\end{split}
\end{equation}

\emph{Example 2.} For the state in \eqref{2.16} with $c_1=c_2=c_3=-c$ and $c\in[0,1]$, the dynamical super quantum discord is given by
\begin{equation}\label{3.7}
\begin{split}
\mathcal{SD}(\tilde{\rho}^{AB})\leq &\frac{1}{d^2}[2(1-c)\log_2{(1-c)}+(1-c+4\gamma c)\log_2{(1-c+4\gamma c)}\\
&+(1+3c-4\gamma c)\log_2{(1+3c-4\gamma c)}-2H(c\tanh{x})].
\end{split}
\end{equation}
The state is a Werner state when $d=2$, and the equality of \eqref{3.7} holds. The difference between super quantum discord and dynamical super quantum discord of the Werner state is given by
\begin{equation}\label{3.8}
\begin{split}
\mathcal{SD}(\rho^{AB})-\mathcal{SD}(\tilde{\rho}^{AB})=&\frac{1}{4}[(1-c)\log_2(1-c)+(1+3c)\log_2(1+3c)\\
&-(1-c+4\gamma c)\log_2{(1-c+4\gamma c)}\\
&-(1+3c-4\gamma c)\log_2{(1+3c-4\gamma c)}].
\end{split}
\end{equation}
Let $T(c,\gamma)=\mathcal{SD}(\rho^{AB})-\mathcal{SD}(\tilde{\rho}^{AB})$. The derivation of $T(c,\gamma)$ with respect to $\gamma$ is
\begin{equation}\label{3.8}
\begin{split}
\frac{\partial T(c,\gamma)}{\partial \gamma}=c\log_2{\frac{1+c-2(2\gamma-1)c}{1-c+2(2\gamma-1)c}}\geq 0.
\end{split}
\end{equation}
Then, $T(c,\gamma)$ is a strictly increasing function with respect to $\gamma$. Thus, $\min{T(c,\gamma)}=T(c,0)=0$ for $c\in[0,1]$. It shows that $\mathcal{SD}(\rho^{AB})\geq\mathcal{SD}(\tilde{\rho}^{AB})$, which implies that the super quantum discord of the Werner state is reduced under the phase damping channel. This is consistent with the results discussed in \cite{JY}.
In addition, according to \cite{HJW}, we can get the quantum discord version of the difference with or without the phase damping channel for the Werner state, i.e.,
\begin{equation}\label{3.8}
\begin{split}
\bar{T}(c,\gamma)&=\mathcal{Q}(\rho^{AB})-\mathcal{Q}(\tilde{\rho}^{AB})\\
&=\frac{1}{4}[(1-c)\log_2(1-c)+(1+3c)\log_2(1+3c)\\
&-(1-c+4\gamma c)\log_2{(1-c+4\gamma c)}\\
&-(1+3c-4\gamma c)\log_2{(1+3c-4\gamma c)}].
\end{split}
\end{equation}
Obviously, $T(c,\gamma) = \bar{T}(c,\gamma)$.
It means that weak measurements capture as much information as projective measurements for the Werner state.
Therefore, super quantum discord and quantum discord are different resources.

\emph{Example 3.} For a state in \eqref{2.16}, when $c_1=0.2, c_2=0.35, c_3=0.1, \gamma=0.9, x=4$, $d=2$, we have $\mathcal{SD}(\rho^{AB})-\mathcal{SD}(\tilde{\rho}^{AB})=-0.0180$. Therefore, we obtain $\mathcal{SD}(\rho^{AB})\leq\mathcal{SD}(\tilde{\rho}^{AB})$.

These models show that usually the super quantum discord of the state in \eqref{2.1} may decrease or increase under the phase damping channel.

\section{Conclusions}
It is known that quantum discord can capture quantum correlations for mixed states that goes beyond the entanglement. Meanwhile the super quantum discord induced by the weak measurement covers all the values between the mutual information and the normal quantum discord. Therefore, the super quantum discord may be potentially a more useful quantum resource and brings new hope for further study of quantum correlations \cite{AA}.

In this paper, we have considered a high-dimensional generalization of the super quantum discord
by embedding the 2-dimensional weak measurements into the high-dimensional system. This is done by joining the 2-dimensional weak measurements by a particular orthogonal complementary subsystem.
We then compute the super quantum discord and find out that some of the distinguished feature of quantum discord are still viable in the high-dimensional case.
We have derived analytic solutions of super quantum discord for two classes of higher-dimensional states and the bounds of the special cases are drawn in graphs. The relationships among the quantum mutual information $\mathcal{I}(\rho^{AB})$, the classical correlation $\mathcal{J}(\rho^{AB})$ and the super quantum discord $\mathcal{SD}(\rho^{AB})$ are discussed. Then, we also compute the dynamical super quantum discord under phase damping channels. We show that the dynamical super quantum discord can be more and less than the usual super quantum discord through examples.
\vskip 1in

\centerline{\bf Acknowledgments}
\medskip
The research is supported in part by the NSFC
grants 12126351, 12126314 and 11871325, and Simons Foundation
grant no. 523868.

\bigskip

\textbf{Data Availability Statement.} All data generated during the study are included in the article.

\bigskip

\bibliographystyle{amsalpha}

\end{document}